# Umbilical Cord Blood Banking and its Therapeutic Uses

Nivethika Sivakumaran[1], Imesha Rashmini Rathnayaka[1], Rashida Shabbir[1], Sasini Sandareka Wimalsinghe[1], J. A. Sumalimina Jayakody[1], Mahisha Chandrasekaran[1]

[1]International College of Business and Technology, Biomedical Science Department, No 502A, R. A. De Mel Mawatha, Sri Lanka

*Corresponding author

Nivethika Sivakumaran

Email: snivethika25@gmail.com

Fax: +94 11 4203170

Author Nivethika Sivakumaran E-mail: snivethika25@gmail.com, Mobile No. - +94771048206

Author Imesha Rashmini Rathnayaka E-mail: imasharashmini@gmail.com, Mobile No. -+94775349616

Author Rashida Shabbir E-mail: rashida7755@gmail.com, mobile No. - +94774319976

Author SasiniSandarekaWimalsingheE-mail: wsasini@yahoo.com, Mobile No. - +94764359634

Author J. A. SumaliminaJayakodyE-mail: sumalimina19@gmail.com, Mobile No. - +94768925901

Author MahishaChandrasekaranE-mail: mahishachandrasekaran@gmail.com, Mobile No. - +94752970463

*Funding Organization: This Literature Review article is funded by International College of Business and Technology,No 502A, R. A. De Mel Mawatha, Bambalapitiya, Colombo 04, Sri Lanka.

**Abstract**

*Umbilical cord blood (UBC) can be viewed as the most promising source of stem cells, in which collection cost is minimal and its benefits are immense. The cord blood is used to treat malignant and nonmalignant diseases; this is due to its progenitor characteristics know as stem cells.*

*Its properties of being, immunologically immature and high plasticity has made it superior to other sources of stem cells. The stem cells collected from cord blood have neutral differentiation capabilities which allow medical professionals to produce functional neural cells from these stem cells.*

*Cord Blood Banking (CBB) is the storing of the umbilical cord blood which is collected immediately after the delivery of the baby. Great care and concern is needed for proper storage of these progenitor cells, hence cord blood banks come into the play, they are of 3 types which are: public, private and direct donation banks.*

*Clinical trials are still at its very early stages having abundances to still be uncovered but results being obtained have demonstrated high potential and more scope towards effective development therapies and treatments for rare disorders.*

**Key words:** umbilical cord blood, germ line layer, progenitor characteristics, cord blood banking, stem cells, pluripotent, therapeutic uses

**Abbreviations:** UCB-umbilical cord blood; CBB- cord blood banking; EPCs- endothelial progenitor cells; MSC- mesenchymal stromal cells; USSC- unrestricted stomatic stem cells; VSEL- very small embryonic-like stem cells; MLPC- multi lineage progenitor cells; DNA- deoxyribose nucleic acid; HES- hydroxgethyl starch; RCB- residual cancer burden; HLA - human leukocyte antigen; NC- natural killers; WBC- white blood cells; NPBI- non protein bound iron; NC- natural cell; GvHD- graft vs host disease; MNCS- mono nucleated cells; RBC- red blood cells; TNC- total nucleated cell; CD133,CD34,CD45- commonly used markers of Hematopoetic progenitor cells and endothelial cell (Cluster of Differentitation; GMO- granulocyte macrophage.

1. **Introduction**

The umbilical cord is thin, long tube-like structure highly composed of muscle that is needed to form a connection between the fetus and the placenta, in the mother's uterus. To provide sufficient and effective circulation of the blood to the growing fetus, the umbilical cord has one vein which carries oxygenated rich blood and two arteries which carry deoxygenated blood. These 3 blood vessels coil around the vein in a helical configuration to form the umbilical cord (Sadler, 2006)

After birth, the blood in the umbilical cord can be collected in two ways, either by a syringe or bag method (Phuc Van Pham, 2014). The cord blood is used as an alternative to bone marrow to restore immunological dysfunctions and for transplantations. This blood is found and collected from the umbilical cord in new born infants (Weiss, 2006).

Present statistics have proved that nearly 80 diseases can be completely cured using umbilical cord blood stem cells and over 50 000 transplants have been successfully carried out worldwide. Some of the diseases cured are cancers and blood disorders which had so far always resulted fatal to many affected (Roura, 2015).

The properties of stem cells remain unknown, but their importance is beyond description. The cells in the umbilical cord are unaffected by the external environment, and can effectively differentiate into many types of cells. Life threatening disorder such as thalassemia can be cured with this stem cell, around the world there are 27 million babies affected with blood disorders which can now be cured with the use of the umbilical cord stem cells. The umbilical cord blood is used to cure disorders and diseases for an individual to whom the umbilical cord belongs to, therefore it is assured that there is no mismatching or rejection by the body, like in many cases, people with rare blood groups find it difficult to find a matching donor. This major problem is solved when umbilical cord blood stem cells are used. There is no counter effect after the treatment. Another important factor to consider is that, if harmful diseases are not cured over generations, the threat is higher and become public concern. Resulting in the wide spreading of the diseases and increase in numbers affected. When such a disease is cured with umbilical cord blood stem cells, the effect of it being passed on is reduced and therefore works towards making the disease endemic (Carroll, 2015).

**2. Extraction and preservation of umbilical cord blood**

There are different methods for both preservation and extraction, but parents are not often given the choice of choosing what they like. Whether it be donating for a government hospital or storing in a private cord blood for personal medicinal purposes.

The sole purpose of processing is to separate stem cells from the cord blood so that a sample is produced that can be used safely (Cord Blood Banking – How Baby's Cells are Extracted & Stored? | LifeCell, 2017).The separation and the processing of UCB samples that are present in a greater volume intended for storage in cord blood banks use a partially automated system to make sure that the large numbers of samples are processed efficiently. A closed system is usually used to reduce the risk of any bacterial infection due to a contamination. This processing method allows the recovery of nucleated cells and progenitors adequately to enable the engraftment procedure(Saunders Comprehensive Veterinary Dictionary, 2016).

Early studies have shown that the use of density gradient techniques to separate cord blood leads to the loss of mononuclear cells. This proves that cord blood should be stored in an unseparated manner. Even though reduction of volume is more economical and efficient. There are several different methods that are used to reduce the volume so as to prevent loss of progenitor cells during cryopreservation (Saunders Comprehensive Veterinary Dictionary, 2016). Such as, density gradient separation that is used for separating particles such as DNA where the sample will be placed on a preformed gradient like sucrose or cesium chloride [7], sedimentation of red cells by gelatin, rouleaux formation induced by hydroxyethyl starch (HES) where the volume of cord blood can be reduced by enhancing sedimentation of red cells by rouleaux formation that is induced by a strong sedimenting agent like 6% hydroxyethyl starch (Hespan), centrifugation, and differential centrifugation with expression of RCB and plasma.

Midwives who are trained in collecting cord blood, covering a period of 24 hours. At the time of delivery; sex, weight and condition of the infant is normally learned. The cord is then being doubly clamped and transacted within 10– 30 seconds after the delivery of newborns.  To ensure sterility the free end of the cord is wiped in betadine(Cord Blood Banking – How Baby's Cells are Extracted & Stored? | LifeCell, 2017).

When the placenta is still inside the uterus, the umbilical vein is punctured, and cord blood is collected using the gravity in the collection bag. If the birth of the baby is caesarean section or it is a multiple birth, the UCB is collected from the placenta that is removed. No matter the birth type, always a maximum effort is taken to obtain the highest possible amount of UCB. From the collected UCB, 20ml of the venous blood that was obtained is frozen at−80°C. The other amount of UCB units that was stored in 4°C be transported, with all the related paperwork and is processed within 24 hrs. The collected amounts of UCB units are then calculated excluding the weight of the bags(Dı´az, 2000).

Several different tests should be run of the collected UCB; therefore, 3ml samples are taken for HLA typing, Nucleated cell (NC) count, CD34+ cell count, Progenitor cell assays, Tests for fungal, aerobic and anaerobic bacteriology cultures. The UCB unit will be transferred to a 150-ml bag and the HES solution will be directly added to the collection bag under sterile conditions which has a proportion of 2:1 of washing the bag before mixing it with the cord blood. The UCB units will be first centrifuged at 40 g for 5 minutes; because of this the WBC (White Blood Cells)-rich supernatant is expressed by the NPBI (Non-Protein Bound Iron)Compomat G4 system into the original collection bag. A second centrifugation will be done at 400 g for 10 minutes where the plasma is discarded into a satellite bag. After these centrifugations, the remaining volume of UCB units has a mean volume of 27 ml of WBC and plasma. The process is performed in a closed system with the use of sterile connecting devices to reduce bacterial infections due to contamination. The NK (Natural Killer) and CD34+ cell estimations are repeated using the pelleted fraction so that there is a monitoring of the quality of the process. WBC amounts present in the collecting bag is transferred to a freezing bag(Dı´az, 2000).

Cryopreservation of the cord blood is done by cryopreservation of processed UCB units in an automated microprocessor-controlled rate freezer. After the WBC is chilled the cold freezing cryopreservative solution containing 60% DMSO is added drop wise for 15 minutes. Samples for quality control of cryopreservation procedure is then extracted before freezing and cryopreserved into cryotubes with the bag. The cells will then be immediately placed inside aluminium cassettes in the chamber of the cell freezer, that uses two thermocouple probes placed in a sample containing the freezing solution (M-Reboredo N, 2000).

The cryopreservation methodology is by 1°C/min cooling down to −60°C, followed by a drop to −120°C, 5°C/min.

At the end of the freezing procedure the cells are stored in liquid nitrogen freezer. Immunophenotyping of the CD34+ cell estimation iscarried out on the whole blood before starting the processing and the volume reduction. The CD34+ cell number is calculated on the amount of the WBC units present after the processing(Dı´az, 2000).

## 3. Advantages of Cord Blood Treatment

The collecting procedure of cord blood is simple and has no medical risk to the mother or newborn baby. The midwives and the responsible staff collects the cord blood from the placenta and then sends it for processing. The cord blood is collected in advance therefore there is plenty of time to test it and for storage and make sure that it is ready use when needed( Advantages and Disadvantages of Cord Blood Treatment, 2017)].

Perhaps one of the most important advantage is that cord blood transplantations do not require a perfect match (HLA typing). Research studies have shown that cord blood transplants can be performed in cases that the donor and the recipient are partially matched. This is because even the partially matched cord blood transplants can be performed where cord blood increases the patient's chance to find a more suitable donor. An estimated number of example, that a national inventory of 150,000 cord blood units that would provide acceptable matches for at least 80-90% of United States patients (Are there and unfavorable aspects of cord blood? - National Cord Blood Program, 2017).

Umbilical stem cell cord blood promises to provide the solution to many critical medical conditions. Most cord blood transplantation has been associated with lower rates of GvHD (Graft vs Host Diseases)(Are there and unfavorable aspects of cord blood? - National Cord Blood Program, 2017).

## 4. Disadvantages of Cord Blood Treatment

The volume of cord blood collected is relatively small therefore the quality stem cells that are used for transplantation much less than that in peripheral blood or bone marrow. If the average total nucleated cell dose (number of nucleated cells per kilogram of the patient's weight) in a cord blood is less than about 1/10th of the average bone marrow then as a consequence, the engraftment (the return of nucleated blood cells, red blood cells and platelets) to the patient's blood is slower with cord blood than with bone marrow transplants. This is a major problem for adults and adolescents because they need more quantity of stem cells for transplantation (Maslova, 2015)].

Cord blood transplantation can exposethe patient to one of the rarest genetic disorders of the immune system or blood. This disorder is not detectable while testing the cord blood sample, but it remains (What are the advantages of cord blood? - National Cord Blood Program, 2017).

The donor cord blood stem cells that are donated by a newborn baby are unavailable for an extra donation of cord blood. Therefore, if by any chance the first cord blood unit fails, then a second unit should be obtained from a different donor.

The American Academy of Pediatrics' says that the chance of a baby needing its own stored cells is approximately 1:1000 to 1:200,000. Therefore, private blood banking is a waste of money and resources ( Advantages and Disadvantages of Cord Blood Treatment, 2017).

## 5. Comparison between old and new methods of extracting and preserving umbilical cord stem cells

### 5.1. Early extraction methods

The collected cells were cultured in a dish and then transplanted into mice, ex vivo expansion and then in vitro transplantation. Once the cells were grown, they were spun in a centrifuge in order to spin them down to separate and extract them. A major limitation observed in this method was that too much plasma contents were collected with not enough MNCs (Mono Nucleated Cells) and no reliable way to concentrate and isolate stem cells(Hussain, 2012).

## 5.2. Modern extraction methods

The modern methods derived by doing the similar steps in an automated way. The cord blood is first separated into several layers such as a layer of RBCs (Red Blood Cells), a layer of plasma and an in between layer which is known as the Buffy layer. Buffy layer is known to be rich with white blood cells and most essential stem cells. Then a suitable processing method is used, to help better separation of cord blood into these multiple layers, allowing for easier extraction of more stem cells (Hussain, 2012).

Five separation methods used are Plasma Depletion, Density Gradient, Hetastarch, PrepaCyte and Automated centrifugal machine (Kawasaki-Oyama, 2008).

Currently the PrepaCyte is the latest and best used proprietary method. It is very similar to the closed method but uses a machine (Basford, 2010).

## 6. Comparison

As it is known umbilical cord blood (UBC) is a rich source of stem cells which can be used to treat diseases. However, earlier, this cord waste discarded as waste material as its properties were hidden. For these stem cells to come into use the umbilical cord blood has to be collected and stored for use when needed. These cells can easily lose their differentiable properties by external conditions if not looked into carefully (Hua, 2013).

Shortly after the delivery of the baby, the blood from the umbilical cord is collected using a syringe and cannula and collected into a bag containing antibiotics and other necessary elements for keeping the blood safe until it is correctly preserved. This is called the closed technique of collection since the umbilical cord is not cut or disturbed in order to collect the blood(Hussain, 2012).

Cryopreservation techniques are then used to store the blood in blood banks – public and private. Cryopreservation is the use of extremely low temperatures maintained in order to preserve the structure of intact living cells, along with creating a stable environment through which the cells can be preserved and stored for future use. This method is considered most easy and reliable (Keiger, 2011).

In order to collect large number of stem cell the extraction process must be entirely or partially automated. This closed system to its maximum extent prevents the possible happenings of bacterial contamination after collection and before storage. It is also needed that maximum number of nucleated cells and progenitors to enable engraftment. This was the disadvantage seen in early separation techniques by density gradient techniques which resulted in the loss of mononuclear cells; hence the modern methods derived which suggested that storing the cord blood before separation. But this too is not the most ideal method as it is essential for minimum volume to be stored in cord blood banks as only then will it be economical and efficient (Beeravolu, 2017).Yet, the modern methods are being updated in order to reduce the volume to be stored along with reducing the loss of progenitor cells. Some modern methods are density gradient separation; sedimentation of red cell by gelatin, rouleaux formation induced by hydroxyethyl starch (HES) and centrifugation, and differential centrifugation with expression of RCB and plasma (M-Reboredo N, 2000).

Finally having an optimal yield of mononuclear cells is the most important aspect of UCB collection. Along with considering the factors which influence the amount of fetal blood which is actually remaining in the placenta and umbilical cord after clamping and dissection (M-Reboredo N, 2000).

From this understanding it is easy to understand the advantages of modern extraction and preservation methods. Firstly, currently the entire segment of the umbilical cord is stored, and this helps to preserve many varied types of cells which can be used in future therapies including cell types such as mesenchymal stem cells, endothelial cells, perivascular cell, growth factors and proteins and epithelial cells which all can be found in parts of the stored segment. It also processes the cord blood after preservation and this helps is treating a wider range of treatments and able to be used for more clinical applications as the stem cells will be more accepted in such cases. Also, cord blood is stored in over 6 vials which helps in samples being thawed independently at different times and also allows the potential for multiple therapeutic uses (Taghizadeh, 2017).

Since UCB is known to have higher frequency of progenitor cells, and higher number of early and committed progenitor cells, their ability to form colony forming units' granulocyte macrophage CFU-GMO is highest. Along with this property they also have non-hematopoietic stem cell and other cell type precursors (Basford, 2010).

Although UCB stem cells are known to have many benefits in transplantations one major disadvantage is the low amount of total nucleated cell (TNC) number able to be collected from a single unit. Factors such as unit size, number of previous pregnancies, age of mother, limited volumes available for collection from each sample, and processing methods used, affect for the total collectable cell count. Hence, in order to actually make cord blood banking a more practical option, more efficient processing method should be considered(Basford, 2010).

**7. Therapeutic uses of umbilical cord blood**

Umbilical cord blood of human is a rich source of hematopoietic stem cells, totipotent cells and pluripotent cells. These cells provide outstanding health treatments in medical industries. The stem cells in the cord blood has the capability to develop into different type of cells so they can produce organ specific tissue in special conditions, so cord blood is termed as regenerative medicine[6]. UCB is a good hematopoietic source and perhaps one of the most important cells that could be derived from UCB is Natural Killer cells. These cells can kill different targets such as cancer or virally-infected cells without any prior activation(Lubin BH, 2007) (Roura, 2015).

Umbilical cord blood is cryopreserved for future use related to most medical conditions and disorders due to its lifesaving properties. It is also used for developing the therapies for incurable diseases. Some of the diseases cured are cancers and blood disorders which are considered to be fatal diseases (Moghul, 2013).

Umbilical cord blood solves many problems in the medical field as there are no counter effects after treatment. It also helps to recover from the harmful diseases which are not cured over generations as when these genetic disorders are cured then they do not pass on to next generation. Diseases such as Alzheimer's, Arthritis, Asthma, Cancer, Diabetes, Heart diseases and Strokes can be completely eradicated from a family history if cured using umbilical cord blood. It is a prevention method from the genetic disorder being passed down to the following generations(Moghul, 2013).

Like cord blood, the cord tissue which is termed as Wharton's jelly is found in the umbilical cord. This Wharton's jelly with Poly-Vinyl Alcohol is useful for mainly treating the skin wounds for humans (Mothersofchange.com, 2012). There is so much difficulty and failure when therapies are used for healing the chronic skin wounds. Examples for some conditions that treated with stem cells are mentioned in the Table 1 (Waller-Wise, 2011).

Table 1: Examples of Conditions Treated with Stem cell Transplants. The table provides a brief understanding about different types of cancers, blood disorder, congenital metabolic disorders and immunodeficiencies which can be treated using stem cell transplants.

| Cancers | Blood disorders | Congenital metabolic disorders | Immunodeficiencies |
| --- | --- | --- | --- |
| Acute lymphocytic leukemia | Sickle-cell anemia | Adrenoleukodystrophy | Adenosine deaminase deficiency |
| Acute myelogenous leukemia | Fanconi's anemia | Gunther's disease | Wiskott-Aldrich's syndrome |
| Chronic myelogenous leukemia | Thalassemia | Gaucher's disease | Duncan's disease |
| Myelodysplastic syndrome | Evan's syndrome | Hurler's syndrome | Ataxia-telangiectasia |
| Neuroblastoma | Congenital cytopeni[1]a | Hunter's syndrome | DiGeorge's syndrome |
| Hodgkin's disease | Aplastic anemia | Krabbe's disease | Myelokathexis |
| Non–Hodgkin's lymphoma | Diamond–Blackfananemia | Sanfilippo's syndrome | Hypogammaglobulinemia |
| Burkitt's lymphoma | Amegakaryocytic thrombocytopenia | Tay-Sachs' disease | SevereCombined immunodeficiency |

Wharton's jelly serves as a good source to heal the wounds as they have the properties like high plasticity, proliferative, differentiation capability, and also low immunogenicity.

This umbilical cord blood has the capacity to differentiate into blood due to the composition of increased count of hemoglobin, hematocrit, leukocytes, reticulocytes, and nucleated (immature) red blood cells with presence of immature white blood cells (G.H. Mamoury, 2003). Therefore, that the cord blood collected from the newborn infants is used as an alternative to bone marrow in transplantations and to restore immunological dysfunctions (Waller-Wise, 2011).

The hematopoietic stem cell is the main composition of the cord blood which has the capability to differentiate into blood and also, they are known to have the antigens CD133, CD34 and CD45 they can be induced during in vitro differentiate in too many linages such as erythroid, megakaryocytic and monocyticetc (Roura, 2015).

Mesenchymal stem cells derived from cord blood has high morphological and also molecular similarities when compared to mesenchymal stem cells derived from the bone marrow. They are being used in medical industry due to their potential of rapid multiplication (Roura, 2015). The mesenchymal cell is used for growing in popularity due to its properties like immunomodulatory, anti-inflammatory and tissue regenerative properties (Lubin BH, 2007).

Neurons cannot be regenerated but mesenchymal stem cells make it possible where it can regenerate or replace the damaged neurons and also increases the myelination of axons. It mainly helps to reduce apoptotic cell death by maintaining homeostasis. During transplantation people suffer due to rejection of the transplanted organ by the body so this MSCs controls the immune cells and prevents the inflammation and rejection caused due to transplantation. MSCs can also replace and repair the blood vessels so they are termed as potential therapeutic remedy for the patients suffering from stroke (Hamad Ali, 2012).

Multipotent non-hematopoietic stem cellsare the final type of stem cells found in the cord blood. These cells can differentiate in cell surface markers and various other cells which represent the three germ layer cells (Roura, 2015).

The umbilical cord blood banking is used mainly in order to treat the genetic diseases caused which cannot be cured by the normal treatments. Many diseases are treated using the cord blood stem cells because these cells have the potential to transform into the differentiated cells needed by treating these cells in specialized cell culture medium so that, it can even develop into a specific organ tissue needed. This is the main reason for the banking of the umbilical cord blood (Hua, 2013).

When there is a disease in the first-born baby then the cord blood obtained from the birth of next baby can help to overcome the disease likewise that cord blood obtained can be also used for the same child if there is any inherited genetic disorder, so the banking of the umbilical cord blood plays an important role. As this cell has the potential to differentiate into non-hematopoietic tissue such as cardiac, neurologic, pancreatic and skin tissues due to the presence of pluripotent stem cells so it can be used to treat the diseases like bone marrow failure, hemoglobinopathies, immunodeficiency, and also the inborn errors of metabolism. This is the main therapeutic use of UCB banking in the clinical or medical field (B. Anthony Armson, 2015).

There is both public and private banking to preserve or to store the umbilical cord blood. Mostly public bank is recommended by the clinicians as there is no fee required for the public cord blood banking and it does not work for any profit. Also, unrelated transplantations are possible to help the people who are actually in need of it. Public cord blood bank is considered to be easily accessed by the public and also involves direct donations of cord blood (CryoCell International, 2017) (Virginia P. Studdert, 2011).

Even though private bank is not mostly recommended it also plays an important role in some aspects. It can be mainly accessed only for family use also it is suggested as a good choice to get rid of hereditary diseases. This is not easily assessable for public so it can be preserved for related transplantations because only the donor has all the rights on their cord blood unit (Wharton's Jelly: Miracle Tissue, 2017).

## 8. Discussion

The umbilical cord blood was discarded as a waste material but today it is considered to be a regenerative medicine in order to produce the organ tissues (Hua, 2013) This cord blood collected from the umbilical cord is composed of mesenchymal stem cells, hematopoietic stem cells and also multipotent non-hematopoietic stem cells which has therapeutic uses as these stem cells are used to treat cancers, blood disorders, congenital metabolic disorders and immunodeficiencies (Roura, 2015). Cord blood also contains some non-hematopoietic stem cells like EPCs (endothelial progenitor cells), MSC (Mesenchymal Stromal Cells), USSC (Unrestricted Somatic Stem Cells),VSEL (Very Small Embryonic-Like stem cells), MLPC (Multi Lineage Progenitor Cells), and neuronal progenitor cells. Mostly the transplantations of bone marrow require surgery to obtain from the donor and also causes rejection but when the transplantation is performed with umbilical cord blood then it prevents the rejection of transplanted organ and also no need of surgery to obtain the cord blood (Moghul, 2013).

Like umbilical cord blood Wharton's jelly is also derived from the umbilical cord blood which is a rich source of mesenchymal stem cells. Wharton's jelly is a gelatinous tissue within the umbilical cord which forms the umbilical cord matrix in order to cover the veins and arteries of umbilical cord. This Wharton's jelly is used in healing of chronic injuries (Zoe, 2016) Cryopreservation is followed to preserve cord blood and Wharton's jelly as it is an easy and reliable method (Basford, 2010).

The collection of the cord blood is the important process in the banking of umbilical cord blood which takes only some minutes. The collection of the cord blood is done during the delivery by clamping the umbilical cord. Once when the umbilical cord is clamped it is wiped with an antiseptic and then a needle is inserted in order to obtain 60 ml of the cord blood early clamping should not be followed as it leads to anemia (Pegg, 2007).

There are two different types of umbilical cord blood bank banks which is very familiar they are public cord blood bank and private cord blood bank. There is also another one as direct donation cord blood bank (Wharton's Jelly: Miracle Tissue, 2017).

The cord blood has a small chance to be used by the donor so it is better to store it in a public bank as it can save a life. In addition, many legal and ethical aspects must be considered in private cord blood banking. It is said to be a biological insurance stored for the future family use as it is regarded as the once in a life time opportunity many of them chose this option(Buatovich, 2016).

USA pioneered in the umbilical cord blood banking. USA has twenty-eight Public cord blood banks and twenty-nine private cord blood banks (Disadvantages of Cord Blood Banking, 2017). UK also implement both Private and Public Cord Blood banking and there are two public Cord blood banks and six Private cord blood banks in UK. There are no any Public cord blood banks in the Egypt all of them are private cord blood banks, there are five private cord blood banks in Egypt which are established by other countries such as UK, Switzerland and India. Whereas there is an Egyptian private cord blood bank named as cell safe bank (Disadvantages of Cord Blood Banking, 2017). South Africa being prominent country in African continent, has three private UCB Banks.Singapore a well-developed country in Asian continent has only one public cord blood banks. India, as a developing country plays a major role in cord blood banking. India has four public cord blood banks and fourteen private cord blood banks (Seah, 2014). Sri Lanka is about to establish a Cord Blood Bank. This will be available at National Blood Centre under the guidance of Dr. VijithGunasekara(Hettiarachchi, 2012).

Only 0.04% of the stem cell used by the same donor and only 0.07% is used for the familial usage such as for siblings. Even though this reduces 50% of the risk of graft versus host diseases private bank cost additional $ 1,374,246 for each year [24]. More than 600000 UCB units have been stored in world wide. But only >30000 transplants have been performed. Even though there are more than 20 million Adult volunteer donors registered in National Marrow Donor program (Karen K. Ballen, 2013).

**Acknowledgements**

International College of Business and Technology (ICBT) No. 36, De Kretser Place, Bambalapitiya, Colombo 04, Sri Lanka and the Department of Biomedical Science at ICBT is gratefully acknowledged for all the Support and Encouragement.